\documentclass[a4paper,fleqn]{cas-sc}

\usepackage[authoryear]{natbib}

\def\tsc#1{\csdef{#1}{\textsc{\lowercase{#1}}\xspace}}
\tsc{WGM}
\tsc{QE}

\begin{document}
\let\WriteBookmarks\relax
\def\floatpagepagefraction{1}
\def\textpagefraction{.001}

\shorttitle{Resolved Studies of Local Group Galaxies Enabled by ELT-SHARP}    

\shortauthors{Wang et al.}  

\title [mode = title]{Unveiling the Origins of Local Group Galaxies with ELT-SHARP: An Overview of Science Cases}  



%

\author[1]{Hai-Feng Wang}[orcid=0000-0001-8459-1036]

\cormark[1]


\ead{haifeng.wang.astro@gmail.com}


\credit{}

\affiliation[1]{
  organization={Local Universe and Time-Domain Astronomy Laboratory, LuoxiaHong Institute of Astronomy, China West Normal University},
  city={Nanchong},
  postcode={637002},
  country={China}
}

\author[2]{Guan-Yu Wang}
\author[1]{Peng Yang}
\author[1]{Guo-Jun Cheng}
\author[3]{Francois Hammer}
\author[4]{Martin Lopez-Corredoira}
\author[1]{Pei-Bin Chen}
\author[1]{Qi-Da Li}
\author[5]{Xiao-Peng Liu}
\author[5]{Yu-Zheng Wang}
\author[6]{Xiao-Yue Zhou}




\affiliation[2]{
  organization={Dipartimento di Fisica e Astronomia ``Galileo Galilei'', Universit\'a degli Studi di Padova},
  addressline={Vicolo Osservatorio 3},
  postcode={I-35122},
  city={Padova},
  country={Italy}
}


\affiliation[3]{
  organization={LIRA, Observatoire de Paris--PSL, CNRS},
  country={France}
}

\affiliation[4]{
  organization={Instituto de Astrof\'{\i}sica de Canarias},
  city={La Laguna},
  state={Tenerife},
  country={Spain}
}

\affiliation[5]{
  organization={Department of Physics, Hebei Normal University},
  city={Shijiazhuang},
  postcode={050024},
  country={China}
}

\affiliation[6]{
  organization={National Astronomical Observatories, Chinese Academy of Sciences},
  city={Beijing},
  postcode={100012},
  country={China}
}

\cortext[1]{Corresponding author}



\begin{abstract}
The $\Lambda$CDM paradigm provides a successful framework for the formation of cosmic structures on large scales, yet persistent tensions might remain on galactic and sub-galactic scales. The Milky Way and the Local Universe offer a uniquely accessible laboratory to probe these discrepancies through detailed chemo-dynamical studies of resolved stellar populations. In the era of large photometric and spectroscopic surveys, understanding disk assembly, dynamical heating, and galactoseismology requires not only wide-field coverage but also high-angular-resolution, near-infrared spectroscopy capable of penetrating dust-obscured regions and resolving crowded stellar environments. We propose a set of science cases that exploits the advanced capabilities of \textsc{SHARP}, a multi-mode near-infrared spectrograph designed for Extremely Large Telescopes and future space facilities such as the Habitable World Observatory. Operating over $0.95$--$2.45\,\mu$m in a single exposure, \textsc{SHARP} enables efficient access to key stellar and gas tracers across the Galactic bulge, disk, halo, and nearby galaxies. The \textsc{NEXUS} mode provides adaptive-optics-assisted multi-object spectroscopy over a $1.2'\times1.2'$ field with spectral resolutions up to $R\sim17{,}000$ for point sources, enabling gas and stellar kinematics. Complementarily, the \textsc{VESPER} multi-IFU mode delivers spatially resolved spectroscopy in crowded and complex regions, allowing the reconstruction of internal kinematics and stellar population gradients on sub-arcsecond scales. By combining \textsc{SHARP} observations with ongoing and upcoming surveys such as \textit{Gaia}, APOGEE, LAMOST, LSST, 4MOST, and ELT--MOSAIC, this project will map six-dimensional phase-space structures of disk and bulge stars, stellar streams, and nearby disk galaxies (e.g. LMC, SMC). The resulting data will enable detailed investigations of bars, spiral arms, warps, flares, and satellite interactions, providing critical constraints on disk evolution, merger history, and dark matter substructure. These efforts will significantly advance our understanding of the dynamical assembly of the Milky Way and the Local Universe, and provide stringent tests of $\Lambda$CDM predictions on the smallest accessible scales.
\end{abstract}




\maketitle
\section{Introduction}\label{}
Understanding the formation and evolution of the Milky Way (MW) and Local Universe disk galaxies remains one of the central goals of modern astrophysics. Challenges remain on galactic and sub-galactic scales, particularly in the context of disk formation, secular evolution, satellite interactions, and the coupling between baryons and dark matter. The Milky Way and the Local Universe provide an unparalleled laboratory in which these issues can be addressed through resolved stellar populations, combining spatial, chemical, and kinematic information \citep{2016ARA&A..54..529B}.
\twocolumn
In the coming decade, large-scale surveys such as \textit{Gaia} \citep{2025arXiv250301533B}, APOGEE (SDSS) \citep{Majewski_2017}, LAMOST \citep{2012RAA....12..735D}, WEAVE \citep{2024MNRAS.530.2688J}, 4MOST \citep{2023Msngr.190...13L}, LSST \citep{2019ApJ...873..111I}, CSST \citep{2025arXiv250704618C}, and ELT-class facilities \citep{2021Msngr.182...33H} will collectively usher in a golden era of Galactic and Local Universe archaeology and seismology. However, a critical gap remains between wide-field surveys and the high-angular-resolution, near-infrared spectroscopic observations required to probe dust-obscured regions, crowded stellar environments, and dynamically complex structures such as the bulge, bar, inner disk, stellar streams, and satellite remnants. The \textsc{SHARP} instrument \citep{2025JATIS..11c5002M, 2024SPIE13096E..5IS}, with its multi-mode design and near-infrared coverage ($0.95$--$2.45\,\mu$m), is uniquely positioned to bridge this gap and enable transformative advances across multiple fronts of Galactic science.

\textsc{SHARP} is a high-performance, multi-mode spectrograph conceived to fully exploit the capabilities of future Extremely Large Telescopes (ELTs) and next-generation space observatories \citep{2025JATIS..11c5002M}. Its design combines two highly complementary observing modes. The \textsc{NEXUS} mode provides adaptive-optics-assisted multi-object spectroscopy over a $1.2'\times1.2'$ field, with spectral resolutions up to $R\sim17{,}000$ for point sources and simultaneous wavelength coverage from $0.95$ to $2.45\,\mu$m. This capability is ideal for precise chemical abundance measurements and stellar kinematics across diverse Galactic environments. The \textsc{VESPER} mode, a multi-integral-field-unit (multi-IFU) system composed of 12 deployable probes, enables spatially resolved spectroscopy over an AO-corrected field of $\sim24''\times70''$, allowing detailed studies of crowded and dynamically complex regions \citep{2024SPIE13096E..5IS}.

Together, these modes make \textsc{SHARP} exceptionally well suited for dissecting the chemo-dynamical structure of the Milky Way and nearby galaxies, particularly in regions where optical surveys are limited by extinction or crowding. By combining \textsc{SHARP} observations with large survey data, we can construct a coherent, multi-dimensional view of Galactic structure and evolution.

One of the primary scientific drivers of this project is to understand the spatial and chemical structure of the Galactic disk and its evolution over cosmic time. Using multiple stellar tracers and population-based models, we aim to characterize the complex density structure of the outer disk across different ages and chemical compositions. Key questions include the amplitude and efficiency of radial migration, the origin of disk asymmetries, and the location and nature of disk truncations \citep{2015BAAA...57..138C, 2017AJ....153...99C}.

The near-infrared sensitivity and high spectral resolution of \textsc{SHARP}/\textsc{NEXUS} enable robust chemical tagging (or Labeling) of stars over a wide range of Galactocentric radii, including the heavily reddened inner disk and bulge \citep{2025ApJ...985...32H}. By combining mono-age and mono-abundance populations, we will investigate the properties of disk flaring and its connection to internal and external perturbations \citep{2012ApJ...753..148B}. Chemical tagging will further allow us to trace the early star formation history of the thin and thick disks, as well as accreted components such as the Gaia--Sausage--Enceladus, providing strong constraints on chemical evolution models and revealing chemically peculiar populations \citep{WangInPrep}.

Stellar kinematics and dynamics encode a fossil record of the Galaxy’s assembly history. With radial velocities and chemical tracers from \textsc{SHARP}, combined with \textit{Gaia} astrometry, we will investigate galactoseismology through detailed analyses of kinematic substructures and phase-space patterns. These include warps, ridges, snails, arches, moving groups \citep{2023AJ....165..110Y}, and bulk motions, whose origins are linked to bars, spiral arms, satellite interactions, and non-equilibrium processes in many resolved galaxies or Milkly Way like galaxies in the sourhern side.

A focus will be the origin of the Galactic warps and its evolution, distinguishing between gravitational and non-gravitational scenarios or both \citep{2020ApJ...897..119W, 2025ApJ...989..213Z}. We will also study the pattern speeds of bars and spiral arms using stars as dynamical tracers, and test the applicability of the virial theorem in non-equilibrium disk models. These observational efforts will be complemented by N-body and test-particle simulations, enabling direct comparisons between data and theory.

Accurate stellar parameters are fundamental to all aspects of Galactic dynamics and archaeology. We will leverage machine learning and data mining techniques, including Random Forests, Neural Networks, Support Vector Regression, and Bayesian methods—to derive precise stellar parameters ($T_{\rm eff}$, $\log g$, [Fe/H], elemental abundances, and distances) from large spectroscopic datasets \citep{2022ApJS..262...20L}. High-quality \textsc{SHARP} spectra will serve as critical training and validation sets, particularly in parameter regimes poorly sampled by existing surveys. This effort is part of an ongoing project to rediscover the Milky Way through machine learning and deep learning approaches.

To interpret the observational results, we will develop chemo-dynamical models and hydrodynamical N-body simulations using tools such as \textsc{galpy} \citep{2015ApJS..216...29B} and \textsc{GIZMO} \citep{2015MNRAS.450...53H}. These simulations will incorporate stars, dark matter, hot gas, and cold gas to model the coupled evolution of the Milky Way, the Sagittarius dwarf galaxy, and the Large Magellanic Cloud (LMC). Observational constraints from variable stars, carbon stars, and RGB populations will be used to refine models of the LMC’s mass, orbit, and dynamical impact, providing a comprehensive picture of the MW--Sgr--LMC system.

Finally, \textsc{SHARP} offers transformative potential for studies of the Galactic bulge, halo accretion history, dwarf galaxies, and globular clusters. The bulge, shaped by bars and secular evolution, hosts multiple stellar populations whose origins remain debated. \textsc{VESPER}’s spatially resolved spectroscopy will allow us to disentangle bulge, bar or long bar, and inner disk populations, quantify the fractions of boxy/peanut/X-shaped components, and investigate the interface between the bulge and disk.

In the halo, precise measurements of stellar streams, escape velocity curves, and satellite kinematics will constrain the Milky Way’s gravitational potential and matter distribution. These studies will shed light on long-standing problems such as the missing satellites and too-big-to-fail issues, as well as the role of tidal stripping and ram-pressure effects in shaping dwarf galaxies in the synergy with radio band.

In summary, by fully exploiting the unique capabilities of \textsc{SHARP} and synergizing with large-scale surveys and advanced simulations, this project will deliver a comprehensive, high-resolution view of the Milky Way and the Local Universe, providing critical tests of galaxy formation theories and advancing our understanding of disk galaxies in a cosmological context.

\section{Science Case: Probing the Galactic Bulge with SHARP}

The Galactic bulge hosts multiple stellar populations exhibiting distinct structural, kinematic, and chemical properties, reflecting a complex formation history involving early gas collapse, mergers, and secular evolution driven by disk \citep{2020AJ....159..270K, 2018ARA&A..56..223B,2016PASA...33...26B,2013ApJ...763...26O,2008A&A...486..177Z}. While metal-rich stars generally exhibit faster rotation and lower velocity dispersion, metal-poor stars display hotter kinematics and are more centrally concentrated \citep{2017A&A...599A..12Z, 2020MNRAS.498.5629D, 2021MNRAS.501.5981L, 2021A&A...656A.156Q}. Orbital classifications, particularly apocenter-based analysis of RR Lyrae stars, have been shown to outperform metallicity in discriminating bulge populations from disk and halo interlopers \citep{2020AJ....159..270K, 2024A&A...687A.312O}. Nevertheless, many questions remain regarding the relative contributions of different progenitors, the processes of secular evolution, the origins of ancient metal-poor populations \citep{2022ApJ...941...45R}, the X-shaped structure \citep{2012ApJ...757L...7L}, and the identification of distinct accreted components such as Kraken and Aurora \citep{2024MNRAS.528.3198B}.

Leveraging the high spatial resolution and near-infrared coverage of \textsc{SHARP}, we can overcome the limitations imposed by crowding and extinction in the central bulge regions. The multi-integral-field capability of \textsc{VESPER} allows for resolved spectroscopy in densely populated fields, enabling simultaneous measurement of stellar kinematics and chemical abundances for hundreds of stars. The \textsc{NEXUS} multi-object mode complements this by providing high-resolution spectroscopy of selected bulge targets across a larger AO-corrected field, capturing both bright giants and faint main-sequence stars. Together, these capabilities allow chemical tagging or tracing, age-dating, and orbital classification across multiple populations.

\begin{figure}
  \centering
  \includegraphics[width=0.5\textwidth]{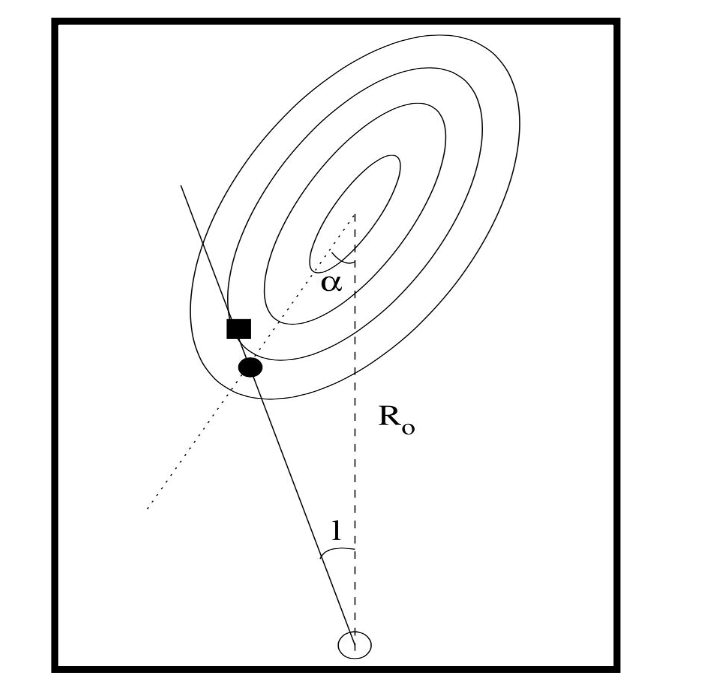}
    \caption{Schematic illustration of the offset between the maximum-density position along the line of sight (squares) in triaxial ellipsoids and the intersection of their major axes with the line of sight \citep{2007AJ....133..154L}. More detailed morphological features, which can only be resolved with high spatial resolution data, are present in \citet{2016A&A...593A..66L}.}\label{bulgefigure}
\end{figure}

Our proposed observations aim to map the chemo-dynamical structure of the inner bulge with unprecedented detail, as seen in Figure~ \ref{bulgefigure}, identifying the fraction of classical versus pseudo-bulge stars, tracing bar-supporting orbits, and characterizing ancient accreted populations. The combination of high spectral resolution, wide near-infrared wavelength coverage ($0.95$--$2.45\,\mu$m), and adaptive-optics-assisted spatial resolution will provide new constraints on formation scenarios, the X-shaped bulge morphology, and the secular evolution of the disk-bulge system. By integrating \textsc{SHARP} data with existing such as OGLE-IV and APOGEE-like atasets, we will deliver a comprehensive chemo-dynamical census of the Galactic bulge, advancing our understanding of its complex formation history and enabling a robust framework for interpreting bulges in external galaxies. 

More specifically, the SHARP spectral range (0.95--2.45\,$\mu$m) provides access or possibilities to a wealth of near-infrared diagnostics. For stellar populations, key absorption features include Mg\,{\sc i} at 1.575\,$\mu$m, Si\,{\sc i} at 1.588\,$\mu$m, Na\,{\sc i} at 2.207\,$\mu$m, Ca\,{\sc i} at 2.263\,$\mu$m, the CO (6-3) band at 1.619\,$\mu$m, and the $^{12}$CO and $^{13}$CO bandheads longward of 2.29\,$\mu$m. These allow robust constraints on $\alpha$-element abundances, iron content, and the cool stellar population properties. Additionally, emission features such as He\,{\sc i} at 1.083\,$\mu$m, Pa\,$\beta$ at 1.282\,$\mu$m, and Br\,$\gamma$ at 2.166\,$\mu$m can probe young stellar objects and active star-forming regions. We therefore anticipate that SHARP will resolve these key diagnostics across the full 0.95--2.45\,$\mu$m window, enabling detailed chemical and dynamical studies of the bulge and nearby galaxies. According to current estimates and in comparison with existing sky surveys, the expected radial velocity accuracy is better than 5--10 km\,s$^{-1}$ --- a precision level adequate for resolving the kinematic signatures of distinct stellar populations. Broadly speaking, VESPER is better optimized for kinematic analyses, whereas NEXUS is more suitable for investigating gas-phase chemical properties.

\section{Science Case: Probing the Spatial-Chemical Structure of Local Universe Disk Galaxies with SHARP}

Understanding the formation and evolution of disk galaxies is a central problem in extragalactic astronomy. Nearby disk galaxies provide an ideal laboratory to investigate the interplay between stellar populations, chemical enrichment, and dynamical processes. The spatial and chemical structure of these disks, including migration, flares, warps, and asymmetries, encodes a galaxy's assembly history, secular evolution, and interaction with its environment. However, resolving individual stellar populations in external disks has historically been limited by crowding and angular resolution (like LMC dwarf galaxy), particularly in dense inner regions. 

\begin{figure}
  \centering
  \includegraphics[width=0.5\textwidth]{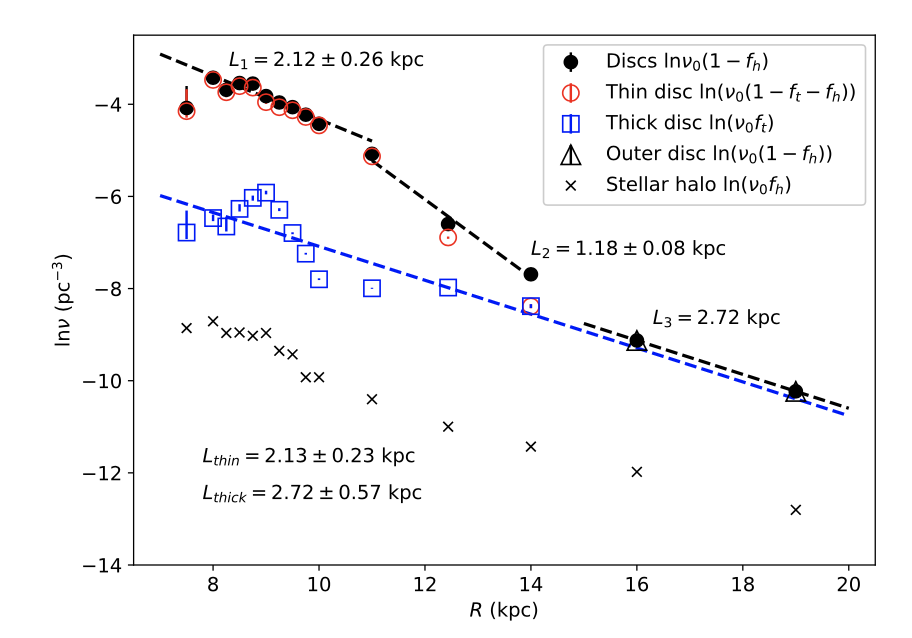}
    \caption{The outer disc exhibits a highly complex spatial structure \citep{2018MNRAS.478.3367W}. Its detailed morphology can only be revealed with high–spatial-resolution observations using large telescopes, such as SHARP.}\label{diskstructurefigure}
\end{figure}

The \textsc{SHARP} multi-mode spectrograph, with its high spatial resolution and wide near-infrared coverage (0.95--2.45~$\mu$m), is uniquely suited to overcome these challenges. Its multi-integral-field unit (\textsc{VESPER}) and multi-object spectroscopy (\textsc{NEXUS}) modes enable simultaneous measurement of stellar kinematics and chemical abundances across dense fields, providing detailed phase-space and chemical maps of stellar populations and gas properties. In nearby disk galaxies such as NGC~891-like, NGC~4013-like, and M31-like in the southern hemisphere, these capabilities allow us to resolve mono-age and mono-abundance populations and trace the radial and vertical structure of the disk with unprecedented clarity. The radial surface brightness profiles of external galaxies are generally classified into three broad types \citep{2006A&A...454..759P}. Understanding which type the Milky Way-like belongs to has long been a topic of interest. In principle, the shape of the radial profile can be used to probe the secular evolution of the Galaxy, as demonstrated by \citet{2008ApJ...675L..65R} and \citet{2006ApJ...645..209D}. More recently, \citet{2018MNRAS.478.3367W} showed that the radial density profile of the Galactic disc exhibits two distinct breaks, shown in Figure~ \ref{diskstructurefigure}.

From a spatial-chemical perspective, we aim to map the outer disk density structure, quantify the amplitude of stellar migration, and identify disk truncations. By applying chemical tracers, we can trace the formation history of thin and thick disk populations, assess the origins of outer disk flares, and distinguish peculiar populations resulting from past accretion events or interactions. High-resolution spectroscopy will allow us to resolve subtle metallicity gradients and abundance ratios that are key diagnostics of star formation histories and gas inflow/outflow processes. Additionally, SHARP’s AO-assisted resolution will mitigate blending effects in crowded regions, improving the accuracy of stellar parameters and chemical abundances.

Meanwhile, the ELT’s large aperture, high spatial resolution, and near-infrared capability will provide unprecedented opportunities to study star formation in the Milky Way and constrain the stellar initial mass function (IMF). Its sensitivity enables the detection of fainter stars, improving sample completeness, extending the low-mass limit, and increasing spatial coverage. The ELT’s angular resolution will significantly reduce biases caused by unresolved binaries, particularly among massive stars. Moreover, its high spectral resolution allows precise measurements of stellar parameters and chemical abundances, enabling robust investigations of IMF variations with metallicity. Near-infrared observations further penetrate dust-obscured star-forming regions, revealing the initial conditions of star formation \citep{2012Sci...337..444S}. The SHARP instrument will enable us to explicitly identify the near-infrared absorption features that we plan to use in this study. For instance, to probe the initial mass function (IMF), we could employ the Na\,{\sc i} doublet or the CO bandheads in the $K$-band. Likewise, we could specify the metallic lines (e.g., Fe\,{\sc i}, Si\,{\sc i}, Mg\,{\sc i}) that will be utilized to trace the ``subtle metallicity gradients.''

From a kinematics-dynamics perspective, we will analyze phase-space structures and identify dynamical signatures such as warps, ridges, snails, and moving groups in the resolved galaxy disks (as least for some nearby dwarf galaxies) with high-angular resolution.At least for some Local Dwarf Galaxies, such studies are feasible with large telescopes, as already demonstrated in previous works (e.g., with HST and JWST). For these targets, SHARP-MOS is sufficient to obtain the required data. In the future, we would be planning to further explore these systems together using SHARP-IFS. The high spectral resolution enables measurement of radial velocities and velocity dispersions, essential for studying the pattern speed of bars and spiral arms and the dynamical heating history of disks. We will combine these observational data with N-body simulations and test-particle models to reconstruct the assembly history and constrain the impact of satellite interactions, bars, and spiral perturbations on disk evolution. Deep learning techniques will further enhance our ability to detect substructures and streams embedded in the disk.

The Milky Way has experienced a rich and complex evolutionary history, and the formation of the Galactic disc has long been a central topic in Galactic archaeology. Although substantial progress has been made, many key issues remain unresolved, including the classification of disc components and their chemical evolution. \citet{2011ApJ...738..187L} suggested that at least two distinct enrichment episodes are required to explain the bimodality observed in $\alpha$-abundance space, often associated with the formation of the thin and thick discs. However, significant overlap between these components exists at intermediate metallicities. In addition, accretion events, such as Gaia--Sausage--Enceladus and the Sagittarius dwarf galaxy, have contributed stars to the disc, further highlighting the complexity of its evolutionary history.

By combining SHARP’s capabilities with multi-band photometry and existing surveys, this project will provide a comprehensive chemo-dynamical census of nearby disks, allowing us to connect stellar populations, structural features, and dynamical processes. The resulting insights will not only advance our understanding of individual galaxies but also inform models of disk galaxy formation and evolution in a cosmological context, bridging the gap between Milky Way studies and extragalactic research.

\section{Science Case: Galactic Halo Dynamics and Accretion History with SHARP}

\begin{figure}
  \centering
  \includegraphics[width=0.5\textwidth]{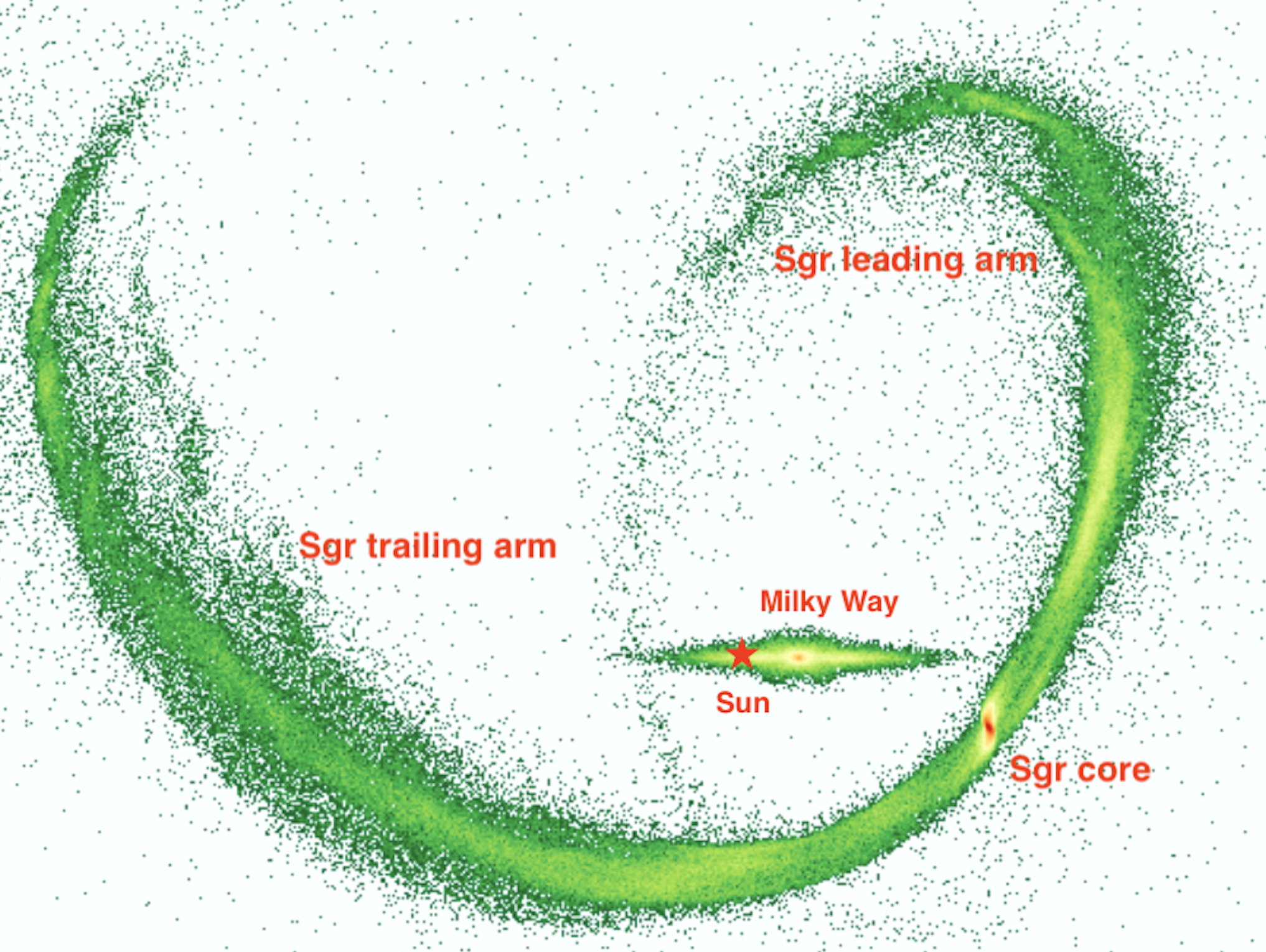}
    \caption{This figure provides an overview of the Sagittarius (Sgr) orbit in the orbital plane, with the Milky Way seen edge-on. The simulation corresponds to our present-day modeling across several hundred runs, clearly showing the trailing arm, leading arm, core, and two possible bifurcations, all labeled in red. The Milky Way disc and the Sun’s location are also indicated. The northern bifurcation is prominently visible, as noted in many previous models, while the most likely southern bifurcation signal is also shown here \citep{2022ApJ...940L...3W}.}\label{sgrfigure}
\end{figure}

The Galactic halo preserves the fossil records of the Milky Way’s assembly history, hosting stellar streams and substructures that trace past accretion events, tidal interactions, and merger processes \citep{2018Natur.563...85H, 2019MNRAS.490.3508L, 2021ApJ...909L..26B, 2019ApJ...872..152I}. Among these, the Sagittarius (Sgr) dwarf galaxy and its extended tidal stream serve as a unique laboratory to study hierarchical galaxy formation, the interplay between satellites and the Milky Way, and the chemo-dynamical evolution of accreted populations \citep{2021ApJ...914..123I, 2022ApJ...926..107M, 2009ApJ...700L..61N, 2012ApJ...753..145C}. However, reproducing the full 6D phase-space properties of Sgr—including its core, leading and trailing arms, which could be found in Figure~ \ref{sgrfigure} , and bifurcations—remains challenging due to the complex gravitational and hydrodynamical interactions within the Milky Way–LMC system \citep{2022ApJ...940L...3W, 2021MNRAS.501.2279V}. The halo stars are typically metal-poor, and many optical features that are prominent in disk stars indeed become weak or disappear in such low-metallicity environments. Nevertheless, in the near-infrared, there remain several proxies. For example, the $H$-band (1.51--1.69\,$\mu$m) offers useful absorption lines such as Mg\,{\sc i}, Si\,{\sc i}, and Ca\,{\sc i}, which have been shown to be accessible even at low metallicities. These features will serve as reliable tracers for the chemical diagnostics we intend to carry out with SHARP.

The \textsc{SHARP} spectrograph, with its high spatial resolution and wide near-infrared coverage (0.95--2.45~$\mu$m), is particularly well-suited to advance halo studies. Its multi-IFU (\textsc{VESPER}) and multi-object (\textsc{NEXUS}) capabilities allow simultaneous measurement of stellar kinematics and chemical abundances across different halo fields, enabling precise mapping of streams and substructures even in crowded regions. By combining these observations with deep learning algorithms, such as our Galactic-Seismology Substructures and Streams Hunter (GS$^3$ Hunter), we can transform high-dimensional phase-space information into robust cluster identification, recovering known streams and discovering new substructures in both the inner and outer halo \citep{2024ApJ...974..219W}.

\begin{figure}
  \centering
  \includegraphics[width=0.5\textwidth]{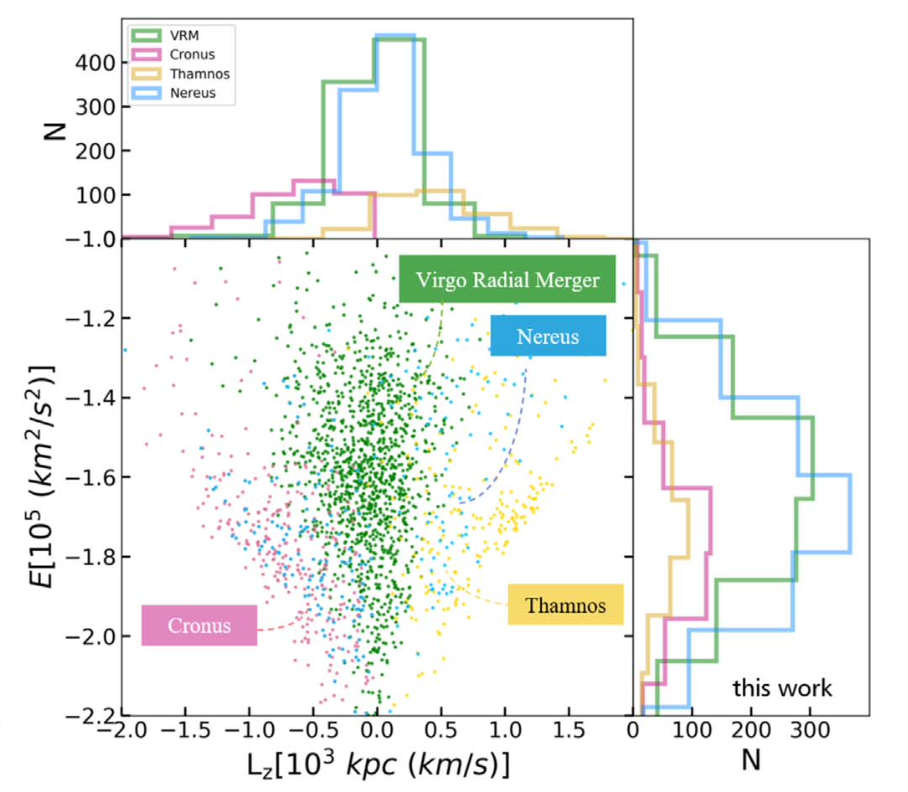}
    \caption{The results obtained with GS3 Hunter are shown, with the components indicated in green (VRM), purple (Cronus), blue (Nereus), and yellow (Thamnos), the figure is from \citet{2024ApJ...974..219W}. To better resolve these structures, future studies should utilize high-resolution observations from instruments like SHARP with enhanced angular resolution.}\label{streamsfigure}
\end{figure}

Our science case focuses on three interconnected aspects: (1) Halo Kinematics and Galactic Seismology: Using SHARP’s AO-assisted resolution and spectroscopic precision, we will resolve the kinematic properties of Sgr-like streams and other halo substructures, enabling detailed studies of orbital dynamics, velocity dispersion (streams in the milky way could be hot and cold), and phase-space morphology. This will allow direct comparison with N-body simulations of the Milky Way–Sgr–LMC system, including variations in satellite mass ratios and the influence of gas dynamics, thereby constraining models of tidal stripping and accretion history. (2) Chemical tracing and Accretion History: High-resolution near-infrared spectra from SHARP will provide accurate elemental abundances for individual stars in streams and the smooth halo. Chemical labelling will differentiate in-situ from accreted populations, elucidating the origins of Gaia-Sausage-Enceladus, Sequoia, Helmi streams, and other halo substructures. These measurements are crucial to reconstruct the evolution history of the Milky Way and its satellites. (3) Discovery and Characterization of new halo substructures ({Figure~ \ref{streamsfigure} labeled by different colors}): By leveraging SHARP’s spatial and spectral resolving power, we aim to extend GS$^3$ Hunter applications to local halo stars, including K giants from LAMOST and upcoming surveys such as LSST, ELT, and China-14.5m. This approach enables identification of previously unknown streams and overdensities, improving our census of the Galactic halo and enhancing our understanding of the Milky Way’s assembly process.

By integrating SHARP’s observational capabilities with deep learning analyses and N-body simulations, this project will provide a high-fidelity chemo-dynamical reconstruction of the Milky Way halo. It will reveal the detailed accretion and tidal stripping history, resolve substructure complexity, and improve constraints on the Milky Way’s mass distribution and rotation curve. The resulting insights will not only advance Galactic archaeology but also provide a benchmark for understanding the formation of stellar halos in other Local Group galaxies, bridging the gap between resolved Milky Way studies and extragalactic galaxy formation models.

\section{Science Case: Dynamics and Chemo-Dynamics of Dwarf Galaxies and Globular Clusters with SHARP}

The halo of the Milky Way hosts a diverse population of stellar systems, including individual stars, stellar streams, globular clusters (GCs), and dwarf galaxies. Among these, dwarf galaxies and GCs, despite possibly having similar stellar masses, provide unique insights into galaxy formation, dark matter distribution, and the dynamical influence of the Milky Way potential \citep{2018MNRAS.478..611B}. The \textsc{SHARP} spectrograph, with its high spatial resolution, broad near-infrared coverage (0.95--2.45~$\mu$m), and integral field unit (IFU) capability, is exceptionally suited to resolve individual stars within these systems, allowing unprecedented studies of their internal kinematics, chemical abundances, and interactions with the Galactic environment. Moreover, SHARP will target key NIR spectral lines to measure abundance variations and identify multiple populations. Specifically, in the $K$-band, we will focus on the Na\,{\sc i} doublet at 2.206 and 2.209\,$\mu$m, as well as the CO bandheads near 2.29--2.34\,$\mu$m, which are sensitive to temperature and surface gravity. In the $H$-band, we will make use of Al\,{\sc i} lines, along with Mg\,{\sc i} and Si\,{\sc i} features, which serve as robust tracers of $\alpha$-element and iron-peak abundances even in metal-poor regimes. These lines have been successfully used in previous near-infrared studies of Galactic globular clusters and dwarf galaxies, and are well within the SHARP spectral coverage (0.95--2.45\,$\mu$m).

\begin{figure}
  \centering
  \includegraphics[width=0.5\textwidth]{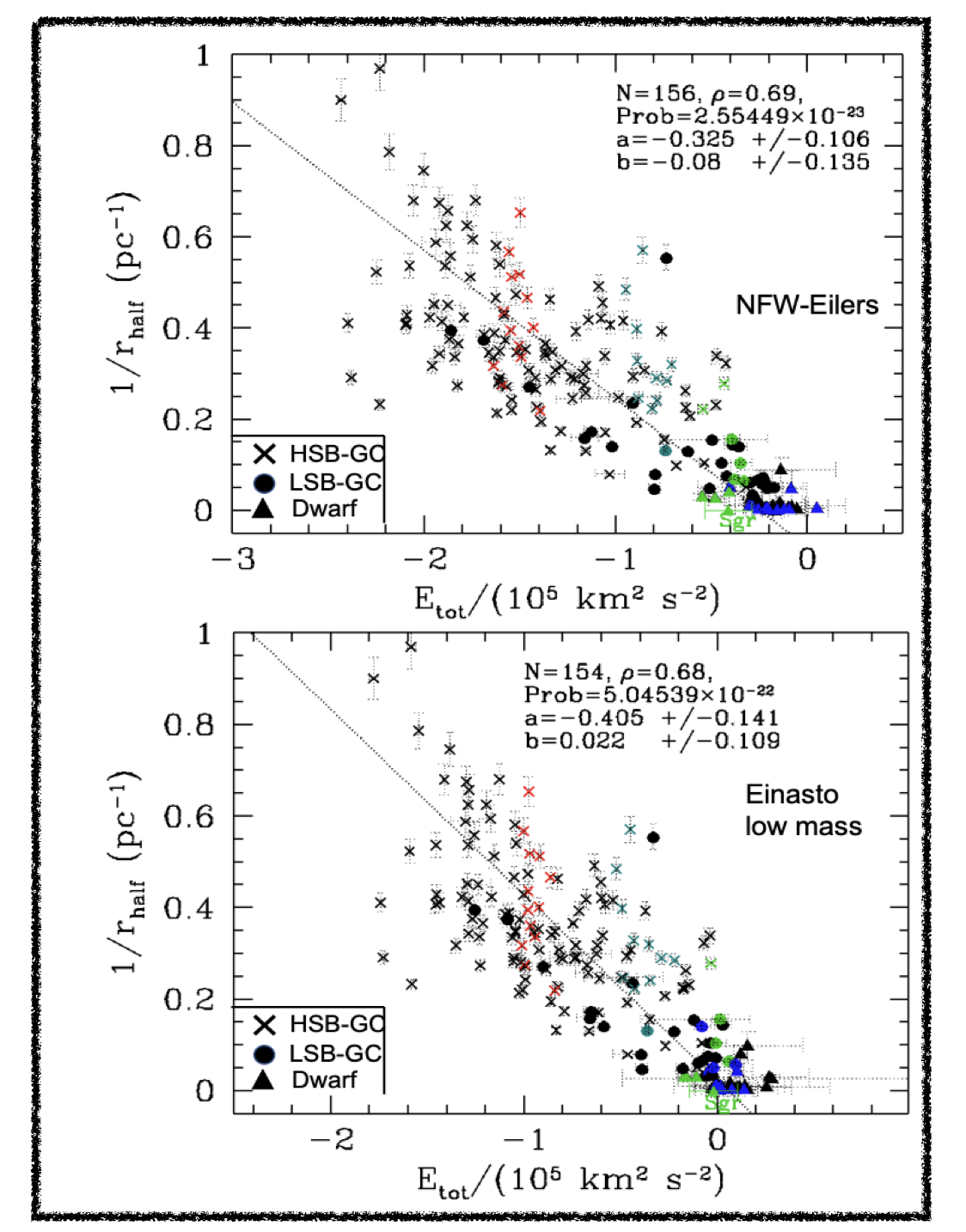}
    \caption{Inverse half-light radius versus total orbital energy for HSB-GCs (crosses), LSB-GCs (filled dots), and dwarf galaxies (triangles). Structures identified by \citet{2022ApJ...926..107M} and \citet{2020MNRAS.498.2472K} are indicated with different color codes (Kraken: red; Gaia–Sausage–Enceladus: cyan; Sgr: green). Dwarf galaxies and LSB-GCs associated with the VPOS \citep{2012MNRAS.423.1109P} are shown in blue. The top (bottom) panel displays 156 (154) GCs and 33 dwarfs, based on total energy calculated from the NFW \citep{1997ApJ...490..493N} (Einasto-low mass\citep{1965TrAlm...5...87E}) model  of the Milky Way. The location of Sgr is marked with a green triangle. The dotted line represents a fit to the GC points, with the corresponding slope (a), intercept (b), and Spearman rank correlation significance given in the top-right of each panel. One-sigma error bars for orbital quantities, such as pericentre and total energy, are derived from Monte Carlo simulations assuming Gaussian errors in proper motion and radial velocity..}\label{dwarfsandgcsfigure}
\end{figure}

Structural and Dynamical Insights: Observations indicate that the half-light radius of GCs is strongly correlated with their orbital pericenter and total orbital energy (e.g., Figure~ \ref{dwarfsandgcsfigure} from \citet{2024MNRAS.527.2718H}), highlighting the significant influence of the Milky Way’s potential on internal cluster structure. By combining SHARP's IFU spatially resolved spectroscopy with kinematic modeling, we can measure gas dynamics and star formation and tidal distortions in both GCs and dwarf galaxies. This enables some tests of equilibrium assumptions: whether these systems follow the Jeans equation, are pseudo-equilibrated under tidal shocks, or are currently experiencing their first pericentric passages, recently losing gas under ram-pressure stripping. High-resolution IFU data allow mapping of internal kinematics in unprecedented detail, providing critical constraints on dark matter distribution and dynamical heating.

Chemical tracers and Star Formation Histories: SHARP’s spectroscopic capabilities facilitate high-precision abundance measurements across individual stars in GCs and dwarf galaxies. Chemical labeling will be employed to identify distinct stellar populations, trace their formation epochs, and correlate chemical substructures with dynamical histories. This is particularly powerful for disentangling the evolutionary histories of halo inhabitants, from ancient systems like Kraken, Gaia-Sausage-Enceladus, and Sagittarius to recently accreted dwarfs associated with the LMC or the vast polar structure.

{Halo Assembly and Environmental Effects: By systematically comparing orbital parameters, structural properties, and chemical signatures across a broad sample of GCs and dwarf galaxies, we aim to calibrate the epoch of infall and reconstruct the assembly history of the Milky Way halo. The combination of high spatial resolution and IFU spectroscopy enables the study of tidal distortions, ram-pressure effects from the Milky Way’s gaseous halo, and dark matter-induced internal kinematics. These analyses will quantify the impact of tidal heating and stripping on structural parameters, such as half-light radii and velocity dispersion profiles, bridging the gap between internal dynamics and global halo evolution, and revealing possible AGN feedback processes, as suggested by \citet{2023RAA....23a5005C}.

Modeling and Synergy with Simulations: Observational data from SHARP will be complemented with N-body and hydrodynamical simulations, exploring the response of GCs and dwarfs to tidal shocks, disk crossings, and halo interactions. By integrating kinematic, chemical, and structural constraints, we will probe the dark matter content of these systems, their orbital equilibrium states, and the interplay between baryonic and dark matter components.

This case will provide a comprehensive chemo-dynamical map of nearby dwarf galaxies and GCs, offering constraints on the Milky Way potential, insights into dark matter distribution, and a refined understanding of halo assembly. The combination of SHARP’s high spatial resolution, near-infrared sensitivity, and IFU capabilities ensures that even crowded systems can be studied in detail, enabling transformational progress in Galactic archaeology, dwarf galaxy evolution, and globular cluster dynamics. Ultimately, these efforts will establish a benchmark dataset for comparative studies of halo systems across the Local Group, fostering a deeper understanding of galaxy formation and evolution in a cosmological context.

\section{Summary and Outlook}

The science cases outlined above demonstrate the transformative potential of \textsc{SHARP} for studies of the Milky Way and the Local Universe. By combining high spatial resolution, broad near-infrared wavelength coverage, high spectral resolution, and IFU capabilities, SHARP enables detailed chemo-dynamical investigations of stellar populations across disks, bulges, halos, dwarf galaxies, globular clusters, and stellar streams. These observations will bridge the gap between resolved stellar astrophysics and galaxy-scale evolution, allowing us to connect internal structures, chemical enrichment histories, and dynamical states to the hierarchical assembly of galaxies.

Building on the science cases outlined above, \textsc{SHARP} provides a uniquely powerful platform to advance studies of the Milky Way and nearby galaxies by combining high spatial resolution, broad near-infrared spectral coverage (0.95--2.45~$\mu$m), and flexible observing modes through NEXUS (MOS) and VESPER (multi-IFU). The key physical parameters enabled by SHARP is suited for spatially resolved stellar population gradients, and internal kinematics of disks, bulges, dwarf galaxies, globular clusters, and stellar streams. These measurements are essential for disentangling secular evolution, accretion history, dynamical heating, and environmental effects, as well as for constraining dark matter distributions on small and intermediate scales. A brief summary of selected science cases is presented in Table~\ref{tab:sharp_science}.

\begin{table}[htbp]
\centering
\caption{Summary of science cases, key near-infrared observables, and suited observing modes for SHARP.}
\label{tab:sharp_science}
\begin{tabular}{p{1.6cm} p{3.2cm} p{2.2cm}}
\toprule
\textbf{Science Case} & \textbf{Key Observables (Lines/Bands)} & \textbf{Suited Mode} \\
\midrule
Milky Way bulge stellar populations & Mg\,{\sc i}, Si\,{\sc i}, Ca\,{\sc i}, Na\,{\sc i} doublet, CO bandheads & Either (MOS, IFS) \\
\hline
Metal-poor halo stars & Mg\,{\sc i}, Si\,{\sc i}, Ca\,{\sc i} (H-band), Al\,{\sc i} & Either (MOS, IFS)) \\
\hline
Multiple populations in GCs/dwarf galaxies & Na\,{\sc i} doublet (K-band), Al\,{\sc i}, CO bandheads & Either (MOS, IFS) \\
\hline
Gas dynamics \& star formation & He\,{\sc i}, Pa\,$\beta$, Br\,$\gamma$ & Either (MOS, IFS) \\
\hline
Gas-phase chemical properties & [O\,{\sc i}], [C\,{\sc i}], H recombination lines & Either (MOS, IFS) \\
\hline
Local dwarf galaxies (crowded fields) & Na\,{\sc i}, CO, Mg\,{\sc i}, Si\,{\sc i}, Ca\,{\sc i} & Either (MOS, IFS) \\
\hline
Chemo-dynamical mapping of bulge/halo & absorption + emission combined & Either (MOS, IFS)) \\
\bottomrule
\end{tabular}
\end{table}

The primary targets include Milky Way bulge and disk populations (especially in highly extincted regions), nearby disk galaxies where stellar components can be partially or fully resolved, Sagittarius-like halo streams and substructures identified by \textit{Gaia}, LSST, DESI, and LAMOST, and dwarf galaxies and globular clusters in the Local Group. These targets are either already available or will be selected from current and forthcoming large surveys, ensuring strong synergy with existing astrometric and photometric datasets. Observationally, NEXUS is best suited for sparse fields, halo streams, and dwarf galaxy members, offering $\sim$30 slitlets over a $1.2'\times1.2'$ AO-corrected field with spectral resolutions up to $R\sim17000$ for point sources. VESPER is optimal for spatially resolved studies of disks, bulges, dwarfs, and clusters, providing multi-IFU spectroscopy over a $24''\times70''$ field at $R\sim3000$--4000. Typical requirements include S/N$\sim30$--50 for robust abundance measurements and S/N$\sim15$--20 for kinematic and membership analyses.

Based on SHARP-ETC estimates, red giant stars with $K\sim20$--21 in nearby galaxies can reach the required S/N within a few hours of integration per pointing \footnote{This is a rough estimation, more details needed.}, while brighter bulge and cluster stars require substantially shorter exposures. The combination of multiplexing and IFU capabilities ensures high observational efficiency and statistically meaningful samples within realistic time allocations. No major departures from the current SHARP design are required for these science goals, although maximized AO performance and, in the future, enhanced multiplexing or field of view would further strengthen disk and halo studies.

In comparison with other facilities, SHARP complements JWST by offering higher spectral resolution and superior capabilities for stellar kinematics and chemical labeling in crowded fields. Relative to ELT instruments such as MICADO and HARMONI, SHARP’s strength lies in its unique combination of MOS and multi-IFU modes over a wide near-infrared wavelength range, enabling both targeted and panoramic studies. Synergies with \textit{Gaia}, LSST, CSST, ALMA, VLT surveys, and other ELT instruments will be crucial, providing astrometry, deep photometry, and gas observations that together with SHARP spectroscopy will deliver a comprehensive picture of galaxy evolution. Overall, SHARP is poised to play a central role in Galactic and Local Universe dynamics and archaeology, transforming our understanding of disk structure, halo assembly, dwarf galaxy evolution, and dark matter in the ELT era.

Looking ahead, SHARP observations, synergistically combined with large photometric and astrometric surveys and state-of-the-art simulations, will provide stringent tests of the $\Lambda$CDM paradigm on small scales. They will refine constraints on dark matter distributions, tidal and ram-pressure processes, and the role of secular versus accretion-driven evolution. Ultimately, SHARP will establish a new benchmark for near-infrared, high-resolution spectroscopy, opening a decisive window onto Galactic or extra-galactic archaeology and seismology and the formation history of disk galaxies in the nearby Universe.












\bibliographystyle{cas-model2-names}

\bibliography{cas-refs}

\bio{}

\endbio


\end{document}